\documentclass[a4paper]{article}
\usepackage{ISCSLP2026}
\usepackage{ifthen}
\usepackage{amsmath}
\usepackage{url}
\usepackage{tikz}
\usetikzlibrary{positioning,arrows.meta,shapes.geometric,fit,calc}

\newboolean{blind}
\setboolean{blind}{false}

\title{Preserving Pretrained Expert Routing for Paralinguistic-Aware ASR:\\ A System Description for Track 1 of the NVVSpeech Challenge 2026}
\name{
	\ifthenelse{\boolean{blind}}{Anonymous to ISCSLP}
	{Guoqiang Hu$^1$, Chen Geng$^2$, Eng Siong Chng$^1$\sthanks{Corresponding author.}}
}
\address{
  \ifthenelse{\boolean{blind}}{Anonymous to ISCSLP}
  {
  	$^1$Nanyang Technological University, Singapore\\
    $^2$Beijing University of Civil Engineering and Architecture, China
  }
}
\email{
	\ifthenelse{\boolean{blind}}{Anonymous to ISCSLP}
	{guoqiang002@e.ntu.edu.sg, 2108540124004@stu.bucea.edu.cn, aseschng@staff.main.ntu.edu.sg}
}

\begin{document}

\maketitle

\begin{abstract}
This paper describes our system for Track 1 (Understanding) of the NVVSpeech Challenge 2026, which requires transcribing Mandarin and English speech while detecting, classifying, and locating 16 types of non-verbal vocalizations (NVVs), each rendered as an inline tag. We fine-tune the full text decoder of the Mixture-of-Experts (MoE) model Qwen3-Omni-30B-A3B on a large bilingual tagged corpus built by (i) unifying seven public NVV corpora under the official 16-tag scheme, (ii) synthesizing 10k Mandarin and 10k English clips per tag with two commercial TTS APIs and two open-source synthesis models, and (iii) crawling in-the-wild audio, with the synthesized and crawled streams filtered by a quality-control pipeline that combines voice activity detection, forced-alignment duration checks, and---for unlabeled audio---captioner-based label verification. Training is designed to preserve pretrained expert routing: we disable the MoE load-balancing loss and introduce Prior Intervention Routing (PIR), a lightweight, identity-initialized, mode-aware intervention on the frozen pretrained router. On our 20,589-utterance held-out set, the system reaches a Track1Score of 78.5 versus 36.4 for the official baseline evaluated on the same set, and the submitted system obtains a final official Track1Score of 49.82, versus 33.32 for the baseline.
\end{abstract}

\section{Introduction}

Human speech carries far more than words: laughter, sighs, coughs, breaths, and other non-verbal vocalizations (NVVs) convey emotion, attitude, feedback, and even physiological state. Most spoken language processing, however, remains transcript-centric: automatic speech recognition (ASR) normalizes NVVs away, and speech synthesizers rarely control them explicitly. The NVVSpeech Challenge 2026~\cite{nvvchallenge}, built on the NVV-SuperBench framework~\cite{nvvsuperbench} and held as an ISCSLP 2026 special session, addresses this gap with two tracks; we participate in Track~1 (Understanding), which asks systems to produce a \emph{tagged transcript} that renders the spoken content of Mandarin or English audio together with all NVVs from a 16-category inventory, each inserted as an inline tag (e.g., \texttt{[laugh]}) at its transcript-relative position. We call this formulation \emph{rich ASR}, following the rich-transcription terminology of~\cite{fishs2}: its output is a \emph{rich transcript}---the verbatim word sequence interleaved with inline tokens for non-lexical vocal events---so one decoding pass yields both content and paralinguistic events; transcripts and corpora without such tags are \emph{plain}.

Decomposing the baseline's self-reported component scores~\cite{nvvbaseline} under the Track~1 metric (Section~\ref{sec:task}) is instructive: Whisper-medium~\cite{whisper} falls 29.95 points short of a perfect Mandarin score, 21.97 of them on the F1 term alone, whereas driving its 16.29\% character error rate (CER) to zero would recover at most 1.63 points. The bottleneck is therefore not transcription accuracy but \emph{NVV event recall and classification}---especially for rare categories---and our system concentrates its resources there.

Prior work on NVV modeling falls into two lines. Audio event detection~\cite{panns,compare2022} treats NVVs as acoustic events but ignores lexical context and cannot place events within a transcript. Paralinguistic-aware ASR~\cite{nvspeech,funaudiollm,qwenaudio} embeds NVVs as inline decodable tokens, matching the Track~1 output format directly. We follow the second line, scaled to a 30B-parameter Mixture-of-Experts (MoE) audio-language model, Qwen3-Omni-30B-A3B~\cite{qwen3omni}. Fine-tuning a pretrained MoE on this skewed dual-objective task (verbatim transcription plus rare tag generation) raises a routing problem: the load-balancing auxiliary loss~\cite{shazeer2017,gshard,switch} pushes routing toward uniformity and erodes pretrained expert specialization, and the two sub-tasks interfere through the shared router. We address both effects explicitly.

Our contributions are:
\begin{itemize}
\item \textbf{Data}: a bilingual tagged corpus combining seven unified public corpora, roughly 320k synthesized clips (10k per tag per language), and web-crawled audio, with a quality-control (QC) pipeline over the synthesized and crawled streams (voice activity detection (VAD), forced-alignment duration gates; captioner verification for unlabeled audio) that lets the corpus grow from unlabeled data.
\item \textbf{Training}: a routing-preserving recipe for MoE fine-tuning---disabling the load-balancing loss (the largest single gain in our ablations, Section~\ref{ssec:ablation}) plus \emph{Prior Intervention Routing} (PIR), a lightweight, mode-aware intervention branch on top of the frozen pretrained router, initialized as an exact identity map.
\item \textbf{Analysis}: per-category F1 highlights, and a discussion of two underspecified label definitions (\texttt{[hum]}, \texttt{[lipsmack]}) whose prevailing usage in public corpora may diverge from the convention assumed at evaluation time.
\end{itemize}

\section{Task and evaluation}
\label{sec:task}

Given a recording, the system outputs the transcript with every NVV inserted as one of 16 inline tags (\emph{breath, sniff, laugh, cry, cough, throat clearing, sneeze, sigh, gasp, snore, yawn, hum, moan, hiss, lipsmack, burp}). Systems are ranked by
\begin{equation}
\label{eq:score}
\begin{split}
\mathrm{Track1Score} ={}& 100\,\big[\, 0.7 F_1 + 0.2\,(1{-}D)\\
&{} + 0.1\,(1{-}\min(E,1)) \,\big],
\end{split}
\end{equation}
averaged over Mandarin and English, where $F_1$ is the event-based micro F1 (an event is correct only if both its category and its transcript-relative position, after normalization and alignment, are correct), $D$ is the multi-event normalized tag distance (mNTD) measuring tag position deviation, and $E$ is the CER (Mandarin) or word error rate (WER; English) of the tagged transcript with tags as atomic tokens. The challenge is open-resource: no official training set is provided, and all external resources must be disclosed (Section~\ref{sec:data}).

\section{Data}
\label{sec:data}

\begin{figure}[t]
  \centering
  \begin{tikzpicture}[
      font=\scriptsize,
      base/.style={draw, rounded corners=2pt, align=center, inner sep=2pt},
      src/.style={base, text width=20.5mm, minimum height=7mm},
      arr/.style={-{Stealth[length=1.8mm]}, black!75, semithick},
      drop/.style={-{Stealth[length=1.5mm]}, red!65, thin},
    ]
    \node[src, draw=teal!55!black,   fill=teal!12]                      (pub) {public NVV corpora\\[-1pt] {\tiny unified to 16 tags}};
    \node[src, draw=blue!55!black,   fill=blue!10,   right=2.5mm of pub] (syn) {speech synthesis\\[-1pt] {\tiny 2 APIs + 2 open models}};
    \node[src, draw=orange!60!black, fill=orange!15, right=2.5mm of syn] (web) {web crawl\\[-1pt] {\tiny rich-ASR pseudo labels}};
    \node[base, draw=black!55, fill=black!6, text width=31mm, minimum height=7.5mm]
      (qc) at ($(syn.south)!0.5!(web.south)+(0,-4.4mm)$)
      {gates 1--2 (synth + crawl)\\[-1pt] {\tiny 1) VAD: max silence $>$ 1\,s\\[-1.5pt] 2) forced-alignment duration}};
    \node[base, draw=orange!60!black, fill=orange!10, text width=21mm, minimum height=6.5mm, below=2.0mm of qc, xshift=7mm]
      (g3) {gate 3 (crawl only)\\[-1pt] {\tiny \mbox{captioner verifies} \mbox{pseudo tags}}};
    \node[cylinder, shape border rotate=90, draw=green!45!black, fill=green!12, aspect=0.12,
          text width=22mm, align=center, inner sep=1.5pt, below=2.0mm of g3, xshift=-7mm] (corpus)
      {bilingual tagged corpus};
    \node[src, draw=black!50, fill=black!6] (plain) at (pub |- g3) {plain ASR ($\sim$20\%)\\[-1pt] {\tiny no tags; \mbox{LibriSpeech}, \mbox{AISHELL}}};
    \node[base, draw=blue!55!black, fill=blue!10, text width=22mm, minimum height=6.5mm, below=2.0mm of corpus]
      (model) {rich-ASR model\\[-1pt] {\tiny full fine-tune}};
    \draw[arr] (syn.south) -- ($(qc.north)+(-6mm,0)$);
    \draw[arr] (web.south) -- ($(qc.north)+(6mm,0)$);
    \draw[arr] ($(qc.south)+(7mm,0)$)  -- node[right, font=\tiny, pos=0.4] {crawl} (g3.north);
    \draw[arr] ($(qc.south)+(-11mm,0)$) -- node[left, font=\tiny, pos=0.35] {synth} ($(corpus.north)+(-6mm,0)$);
    \draw[arr] (g3.south) -- ($(corpus.north)+(5mm,0)$);
    \draw[arr] (pub.west) -- ++(-3mm,0) |- ($(corpus.west)+(0,-1.6mm)$);
    \draw[arr] (plain.south) |- node[pos=0.85, above, font=\tiny] {anti-hallucination} ($(corpus.west)+(0,1.6mm)$);
    \draw[arr] (corpus) -- (model);
    \draw[arr, dashed, orange!60!black] (model.east) -- ($(web.east |- model.east)+(3mm,0)$) -- ($(web.east)+(3mm,0)$) -- (web.east);
    \node[font=\tiny, text=orange!60!black, anchor=east, align=right] at ($(web.east |- corpus.east)+(2.4mm,0)$) {self-\\[-2pt]training};
    \draw[drop] ($(qc.south east)+(0,1.2mm)$) -- ++(2.6mm,-1.8mm) node[below=-2.5pt, red!65, font=\tiny] {$\times$};
    \draw[drop] ($(g3.south east)+(0,1.2mm)$) -- ++(2.6mm,-1.8mm) node[below=-2.5pt, red!65, font=\tiny] {$\times$};
  \end{tikzpicture}
  \caption{Data pipeline. Synthesized and crawled streams pass VAD and duration gates; captioner verification applies only to the crawled stream; plain (untagged) ASR data stabilizes the pseudo-labeler.}
  \label{fig:data}
\end{figure}
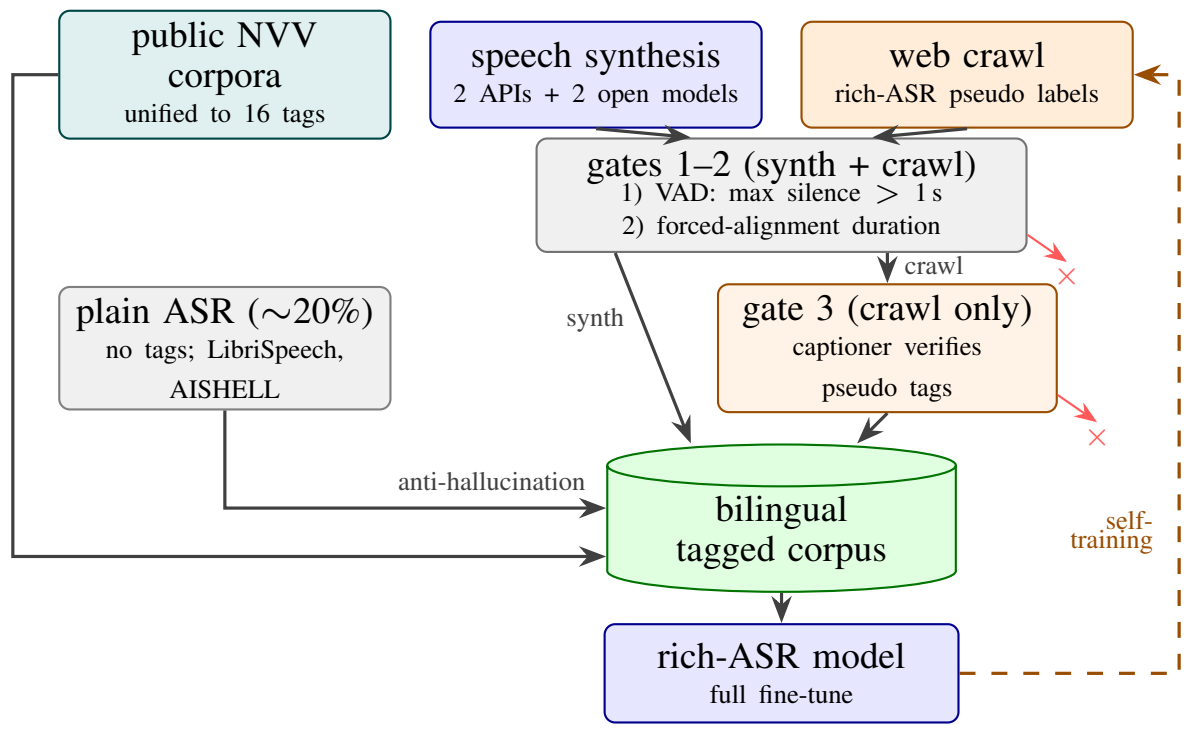

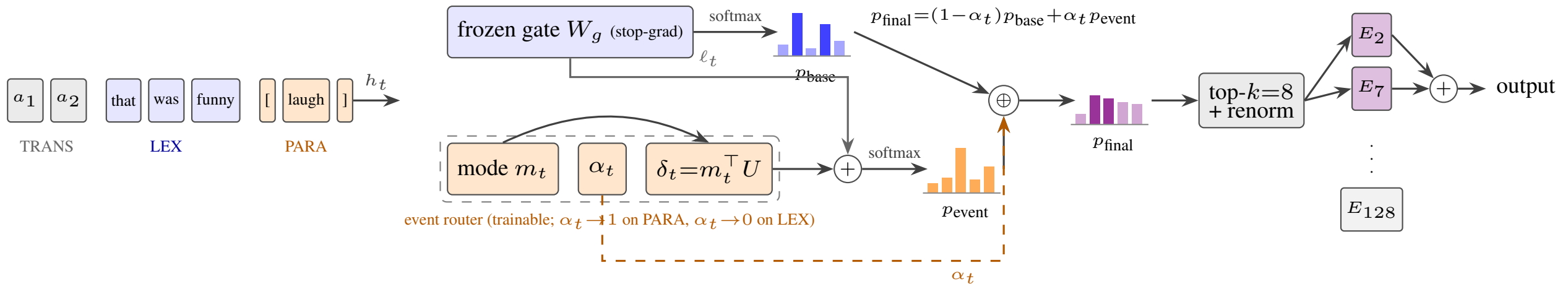
\begin{figure*}[t]
  \centering
  \begin{tikzpicture}[
      font=\scriptsize,
      box/.style={draw=black!70, rounded corners=1.5pt, inner sep=3pt, minimum height=5.5mm, align=center},
      frozen/.style={box, fill=blue!10},
      trainbox/.style={box, fill=orange!20},
      neutral/.style={box, fill=black!8},
      tok/.style={draw=black!70, rounded corners=1pt, minimum height=4.6mm, inner sep=1.5pt, align=center, font=\tiny},
      arr/.style={-{Stealth[length=1.8mm]}, black!75, semithick},
    ]
    \node[tok, fill=black!8]                     (t1) {$a_1$};
    \node[tok, fill=black!8,  right=0.7mm of t1] (t2) {$a_2$};
    \node[tok, fill=blue!10,  right=2.0mm of t2] (t3) {that};
    \node[tok, fill=blue!10,  right=0.7mm of t3] (t4) {was};
    \node[tok, fill=blue!10,  right=0.7mm of t4] (t5) {funny};
    \node[tok, fill=orange!20, right=2.0mm of t5] (t6) {[};
    \node[tok, fill=orange!20, right=0.7mm of t6] (t7) {laugh};
    \node[tok, fill=orange!20, right=0.7mm of t7] (t8) {]};
    \node[font=\tiny, text=black!60, below=0.8mm of $(t1.south)!0.5!(t2.south)$] {TRANS};
    \node[font=\tiny, text=blue!60!black, below=0.8mm of t4.south] {LEX};
    \node[font=\tiny, text=orange!70!black, below=0.8mm of t7.south] {PARA};
    \coordinate (strip east) at (t8.east);
    \coordinate (split) at ($(strip east)+(5mm,0)$);
    \draw[arr] (strip east) -- node[above, font=\tiny] {$h_t$} (split);
    \node[frozen, above right=4.5mm and 5mm of split] (gate) {frozen gate $W_g$ {\tiny(stop-grad)}};
    \node[minimum width=8mm, minimum height=5mm, inner sep=0, right=9mm of gate] (pbase) {};
    \foreach \i/\h/\c in {0/1.2/blue!35, 1/4.6/blue!75, 2/0.8/blue!35, 3/3.4/blue!75, 4/1.6/blue!35}
      \fill[\c] ($(pbase.south west)+(\i*1.5mm,0)$) rectangle ++(1.1mm,\h mm);
    \draw[black!45, line width=0.4pt] ($(pbase.south west)+(-0.5mm,0)$) -- ++(8.3mm,0);
    \node[font=\tiny, below=0.3mm of pbase.south] {$p_{\text{base}}$};
    \draw[arr] (gate) -- node[above, font=\tiny] {softmax} (pbase.west);
    \node[trainbox, below right=4.5mm and 5mm of split] (mode) {mode $m_t$};
    \node[trainbox, right=2mm of mode] (alpha) {$\alpha_t$};
    \node[trainbox, right=2mm of alpha] (delta) {$\delta_t{=}m_t^{\top}U$};
    \node[draw=black!50, dashed, rounded corners=2pt, inner sep=2pt, fit=(mode)(alpha)(delta),
          label={[font=\tiny, text=orange!70!black]below:event router (trainable; $\alpha_t{\to}1$ on PARA, $\alpha_t{\to}0$ on LEX)}] (er) {};
    \draw[arr] (mode.north) to[bend left=25] (delta.north);
    \node[circle, draw=black!70, inner sep=0.6pt, right=6.5mm of delta, font=\tiny] (add) {$+$};
    \draw[arr] (delta) -- (add);
    \draw[arr, black!60] (gate.south) -- ++(0,-2mm) -| node[pos=0.25, above, font=\tiny] {$\ell_t$} (add.north);
    \node[minimum width=8mm, minimum height=5mm, inner sep=0, right=7mm of add] (pevent) {};
    \foreach \i/\h in {0/1.0, 1/1.6, 2/4.8, 3/1.4, 4/2.8}
      \fill[orange!65] ($(pevent.south west)+(\i*1.5mm,0)$) rectangle ++(1.1mm,\h mm);
    \draw[black!45, line width=0.4pt] ($(pevent.south west)+(-0.5mm,0)$) -- ++(8.3mm,0);
    \node[font=\tiny, below=0.3mm of pevent.south] {$p_{\text{event}}$};
    \draw[arr] (add) -- node[above, font=\tiny] {softmax} (pevent.west);
    \coordinate (midy) at ($(pbase.east)!0.5!(pevent.east)$);
    \node[circle, draw=black!70, inner sep=0.8pt, font=\tiny] (mix) at ($(midy)+(8mm,0)$) {$\oplus$};
    \draw[arr] (pbase.east) -- (mix);
    \draw[arr] (pevent.east) -- (mix);
    \draw[arr, dashed, orange!70!black] (alpha.south) -- ++(0,-7mm) -| node[pos=0.45, below, font=\tiny] {$\alpha_t$} (mix.south);
    \node[minimum width=8mm, minimum height=5mm, inner sep=0, right=6mm of mix] (pfin) {};
    \foreach \i/\h/\c in {0/1.1/violet!35, 1/3.1/violet!80, 2/2.8/violet!80, 3/2.4/violet!35, 4/2.2/violet!35}
      \fill[\c] ($(pfin.south west)+(\i*1.5mm,0)$) rectangle ++(1.1mm,\h mm);
    \draw[black!45, line width=0.4pt] ($(pfin.south west)+(-0.5mm,0)$) -- ++(8.3mm,0);
    \node[font=\tiny, below=0.3mm of pfin.south] {$p_{\text{final}}$};
    \draw[arr] (mix) -- (pfin.west);
    \node[font=\tiny, align=center] at ($(mix)+(0,9mm)$) {$p_{\text{final}}{=}(1{-}\alpha_t)p_{\text{base}}{+}\alpha_t p_{\text{event}}$};
    \node[neutral, right=5mm of pfin] (topk) {top-$k{=}8$\\[-2pt] + renorm};
    \draw[arr] (pfin.east) -- (topk);
    \node[tok, fill=violet!25, right=5mm of topk, yshift=7mm]  (e1) {$E_2$};
    \node[tok, fill=violet!25, below=1mm of e1] (e2) {$E_7$};
    \node[font=\tiny, below=0.4mm of e2] (edots) {$\vdots$};
    \node[tok, fill=black!5, below=0.4mm of edots] (e3) {$E_{128}$};
    \draw[arr] (topk.east) -- (e1.west);
    \draw[arr] (topk.east) -- (e2.west);
    \node[circle, draw=black!70, inner sep=0.6pt, right=4mm of e2, font=\tiny] (osum) {$+$};
    \draw[arr] (e1.east) -- (osum);
    \draw[arr] (e2.east) -- (osum);
    \node[right=3mm of osum, font=\scriptsize] (out) {output};
    \draw[arr] (osum) -- (out);
  \end{tikzpicture}
  \caption{Prior Intervention Routing (PIR) in one MoE layer. The frozen gate gives the prior $p_{\text{base}}$; a lightweight event router predicts mode $m_t$, strength $\alpha_t$, and offset $\delta_t$, giving $p_{\text{event}}$; the two are interpolated in probability space before top-$k$ selection. Darker bars mark the (schematic) selected set, whose composition the intervention changes; at initialization ($U{=}0$, $\alpha_t{\approx}0.02$), PIR is an exact no-op.}
  \label{fig:pir}
\end{figure*}

\subsection{Public corpora and label unification}

Figure~\ref{fig:data} overviews the pipeline. We collect the public NVV corpora recommended by the organizers plus the synthetic burp set released with the official baseline~\cite{nvvbaseline}---eight sources, about 298k utterances (848\,h) in total. One of them, WESR-Bench~\cite{wesrbench}, is discarded because its annotations cannot be reliably unified into point-level 16-tag transcripts; the seven remaining sources are Emilia-NV~\cite{nvspeech,emilia}, SynParaSpeech~\cite{synparaspeech}, NonVerbalSpeech-38K~\cite{nvs38k}, SMIIP-NV~\cite{smiipnv}, NonverbalTTS~\cite{nonverbaltts}, MNV-17~\cite{mnv17}, and the burp set. Their taxonomies conflict: the same phenomenon appears as \emph{laughing/laughter/chuckle/giggle}, and several corpora include events outside the official inventory (lexicalized fillers such as \emph{uhm}, or non-vocal events such as \emph{clapping}). We manually map every source label onto the 16 official tags, merging synonyms (\emph{breathing/inhale/exhale} $\rightarrow$ \texttt{[breath]}) and discarding out-of-inventory events. After unification, 206,028 utterances remain; 10\% per tag form a stratified held-out set (20,589 utterances), leaving 183,585 training and 1,854 validation utterances ($\approx$513\,h; Table~\ref{tab:data}). The corpus is heavily long-tailed (85k \texttt{[breath]} vs.\ fewer than 100 \texttt{[snore]} events) and only $\sim$5\% English, motivating the synthesis below.

\begin{table}[t]
  \caption{Public-corpus training pool (after label unification and removal of the 10\% stratified held-out split): 183,585 training + 1,854 validation utterances. zh = Mandarin, en = English.}
  \label{tab:data}
  \centering
  \small
  \begin{tabular}{l r l}
    \hline
    \textbf{Corpus} & \textbf{\#Utts} & \textbf{Notes} \\
    \hline
    Emilia-NV~\cite{nvspeech} & 90,350 & zh, auto-labeled \\
    SynParaSpeech~\cite{synparaspeech} & 36,237 & zh, synthetic \\
    NonVerbalSpeech-38K~\cite{nvs38k} & 34,916 & mostly zh \\
    SMIIP-NV~\cite{smiipnv} & 16,429 & zh, acted \\
    NonverbalTTS~\cite{nonverbaltts} & 4,300 & en \\
    MNV-17~\cite{mnv17} & 1,857 & zh, acted \\
    Baseline burp set~\cite{nvvbaseline} & 1,350 & synthetic \\
    \hline
    Total & 185,439 & $\approx$513\,h (train) \\
    \hline
  \end{tabular}
\end{table}

\subsection{Synthesized data}

To repair the long tail and the language imbalance, we synthesize tagged speech with two commercial TTS APIs---Gemini 3.1 Flash TTS~\cite{gemini} and ElevenLabs~\cite{elevenlabs}---and two open-source models---Maya1~\cite{maya1} and Step-Audio-EditX~\cite{stepaudioeditx}. Carrier texts are generated by GPT-4o~\cite{gpt4o} with explicit constraints on topic and length diversity; tags are then inserted in a second step, using a prompt adapted from ElevenLabs' published audio-tag insertion prompt~\cite{elevenlabs}. For \emph{each} of the 16 tags we synthesize 10k Mandarin and 10k English clips ($\approx$320k clips in total), giving every tag--language pair substantial coverage.

\subsection{Web-crawled data}

We further crawl unlabeled in-the-wild audio from publicly accessible internet sources: recordings are segmented and speaker-attributed with Gemini 3.1 Pro~\cite{gemini}, timestamps are refined with a Qwen3-based forced aligner~\cite{qwen3omni}, and each segment is pseudo-labeled by a rich-ASR checkpoint trained on the labeled portion of the corpus. Pseudo-labeled segments must pass the QC gates of Section~\ref{ssec:qc}---including captioner verification, applied \emph{only} to this unlabeled stream---before entering training, so the labeled pool grows from unlabeled audio in a self-training fashion. To keep the pseudo-labeler well calibrated on ordinary speech and to suppress tag hallucination, we also mix \emph{plain} ASR data (ordinary speech--transcript pairs with no NVV tags) from LibriSpeech~\cite{librispeech} and AISHELL~\cite{aishell} into training; this plain data accounts for roughly 20\% of the total training data.

\subsection{Quality control}
\label{ssec:qc}

All synthesized and crawled data must pass gates 1--2; the unlabeled crawled stream must additionally pass gate 3:
\begin{enumerate}
\item \textbf{VAD} (Silero VAD~\cite{silerovad}): clips whose longest internal silence exceeds 1\,s are discarded (synthesis occasionally yields silent or truncated events);
\item \textbf{Forced-alignment duration check}: a Qwen3-based forced aligner localizes the NVV between the end of the preceding character (or word, for English) and the start of the following one, yielding a maximum-probability time span for the event; clips whose event duration falls outside a preset range are discarded;
\item \textbf{Captioner verification} (unlabeled data only): Qwen3-Omni-Captioner~\cite{qwen3omni} is prompted to list, in temporal order, the NVV events actually produced by the speaker; a pseudo-labeled segment is kept only if the captioner's event sequence is consistent with the pseudo tags. Because the captioner's accuracy and coverage are limited, we do not gate synthesized data (whose intended tags are known) on it.
\end{enumerate}

\noindent\textbf{External-resource disclosure.} Per challenge rules, our system uses the following external resources: the eight public corpora above (including WESR-Bench~\cite{wesrbench}, which we collected but excluded during unification); ESC-50~\cite{esc50}, used only in a negative-result augmentation experiment (Section~\ref{ssec:ablation}); LibriSpeech~\cite{librispeech} and AISHELL~\cite{aishell} (untagged ASR data); the self-collected web-crawled audio above (QC-filtered before use); Qwen3-Omni-30B-A3B and Qwen3-Omni-Captioner~\cite{qwen3omni}; a forced aligner built on Qwen3~\cite{qwen3omni}; Silero VAD~\cite{silerovad}; Gemini 3.1 Flash TTS and Gemini 3.1 Pro~\cite{gemini}; ElevenLabs~\cite{elevenlabs}; Maya1~\cite{maya1}; Step-Audio-EditX~\cite{stepaudioeditx}; GPT-4o~\cite{gpt4o} for carrier-text generation; and Megatron-based fine-tuning infrastructure~\cite{msswift,megatron}. No official evaluation audio, labels, or annotations were used for training or model selection.

\section{System}

\subsection{Backbone and task format}

Our backbone is Qwen3-Omni-30B-A3B~\cite{qwen3omni}, an audio-language model whose text decoder is a 48-layer MoE with $E{=}128$ experts per layer, top-$k{=}8$ routing, and $\sim$3B active parameters per token. We fine-tune the full text MoE---except the pretrained router gates, which PIR freezes (Section~\ref{ssec:pir})---while keeping the audio encoder and the audio--text aligner frozen. The task is cast as instruction-following ASR: a fixed system prompt (in Chinese) instructs the model to transcribe verbatim and insert bracketed tags from the 16-tag inventory at the corresponding positions; the target is the tagged transcript. Tags are written as plain text and tokenized by the ordinary byte-pair-encoding (BPE) tokenizer; this avoids the official baseline's vocabulary hijacking---no embedding surgery is needed---and tag tokens keep their pretrained textual semantics.

\subsection{Disabling the load-balancing loss}
\label{ssec:aux}

MoE pretraining employs a load-balancing auxiliary loss~\cite{gshard,switch} to spread tokens uniformly over experts. In downstream supervised fine-tuning (SFT) on a long-tailed task, this loss is actively harmful: it keeps pushing routing toward uniformity, overriding the pretrained expert specialization that rare NVV tokens depend on. We therefore set the auxiliary-loss coefficient to zero for all fine-tuning; in the ablation of Section~\ref{ssec:ablation}, this single change lifts tagging F1 on a rare-class-focused subset from 0.652 to 0.822---the largest single gain in our system.

\subsection{Prior Intervention Routing (PIR)}
\label{ssec:pir}

Even with the balancing loss disabled, transcription and tag generation share one router, and full fine-tuning lets the task gradient overwrite pretrained routing priors; PIR mitigates both by \emph{freezing} the pretrained gate and learning a small, mode-aware \emph{intervention} on top of it (Figure~\ref{fig:pir}).

Let $h_t \in \mathbb{R}^{d}$ be the hidden state entering an MoE layer, and let the frozen pretrained gate $W_g$ produce logits and prior probabilities
\begin{equation}
\ell_t = W_g h_t, \qquad p_{\text{base},t} = \mathrm{softmax}(\ell_t).
\end{equation}
An \emph{event router}, added in parallel, consists of three lightweight heads. A \textbf{mode head} classifies each token into $M{=}3$ modes---LEX (lexical transcript tokens), PARA (tokens inside a bracketed tag span), and TRANS (transcription context: acoustic, prompt, and structural tokens):
\begin{equation}
m_t = \mathrm{softmax}(W_m h_t + b_m) \in \Delta^{M-1}.
\end{equation}
An \textbf{alpha head} outputs an intervention strength $\alpha_t \in (0,1)$, and a \textbf{delta head} produces a mode-conditioned per-expert logit offset $\delta_t$ from a learnable matrix $U \in \mathbb{R}^{M\times E}$:
\begin{equation}
\alpha_t = \sigma(w_\alpha^{\top} h_t + b_\alpha), \qquad \delta_t = m_t^{\top} U.
\end{equation}
The final routing distribution interpolates the prior and the intervened distribution \emph{in probability space}:
\begin{equation}
\begin{split}
p_{\text{event},t} &= \mathrm{softmax}(\ell_t + \delta_t),\\
p_{\text{final},t} &= (1-\alpha_t)\, p_{\text{base},t} + \alpha_t\, p_{\text{event},t},
\end{split}
\end{equation}
followed by standard top-$k$ selection and renormalization on $p_{\text{final},t}$. Because the mixture need not preserve the ranking induced by $\ell_t$, the intervention can change the \emph{set} of selected experts---precisely the source of its expressive power.

\textbf{Mode supervision.} Mode labels are derived from the target sequence at zero cost: tokens inside \texttt{[}\dots\texttt{]} spans are PARA; other assistant tokens are LEX; everything else is TRANS. A weighted cross-entropy $\mathcal{L}_{\text{mode}}$ (class weights LEX:PARA:TRANS $=1{:}2{:}0.5$) supervises only the mode head at every MoE layer; the total loss is
\begin{equation}
\mathcal{L} = \mathcal{L}_{\text{ASR}} + \lambda \sum_{\text{layers}} \mathcal{L}_{\text{mode}}, \qquad \lambda = 0.01.
\end{equation}
Unlike the load-balancing loss, this signal never touches the frozen gate.

\textbf{Identity initialization.} We set $U{=}0$, $W_m{=}0$, $w_\alpha{=}0$, and $b_\alpha{=}-4$ (so $\alpha \approx 0.018$), making PIR an \emph{exact no-op} at initialization: training starts from the unperturbed pretrained router (the base gate receives no gradient) and learns when and how to intervene. The added parameters ($\sim$8.6k per layer, $\sim$0.4M total) are negligible relative to the 30B backbone, and inference uses the same routing computation with predicted modes, so training and inference routing match exactly.

\subsection{Data rebalancing}
\label{ssec:rebal}

To counter the long tail, we downsample breath-only utterances by half and upsample the seven rarest tags $5\times$ and English data $2\times$ (161,978 public-corpus training utterances); Section~\ref{ssec:ablation} shows this stacks with disabling the balancing loss (Section~\ref{ssec:aux}).

\subsection{Inference details}
\label{ssec:infer}

We decode greedily (temperature 0, repetition penalty 1.1, at most 512 new tokens). Post-processing removes bracketed spans outside the 16-tag inventory, truncates degenerate $n$-gram repetition runs (cap 12; longest legitimate run: 10), and falls back to \texttt{[breath]} for empty outputs; this deterministic, rule-based cleanup applies identically to all evaluations, including the official submission, with no manual edits. Held-out inference (20,589 utterances) takes about 40 minutes on 8 GPUs.

\section{Experiments}

\subsection{Setup}

All models are fine-tuned with Megatron-SWIFT~\cite{msswift,megatron} on 8$\times$A100-80GB GPUs (tensor parallelism 2, expert parallelism 8, sequence parallelism, bf16): Adam ($\beta_1{=}0.9$, $\beta_2{=}0.95$), lr $10^{-5}$ with cosine decay to $10^{-6}$, 2\% warmup, weight decay 0.1, gradient clipping 1.0, global batch 64, max sequence length 4,096, one epoch over the respective training mix. Unless stated otherwise, results are on our 20,589-utterance stratified held-out set (19,619 zh / 970 en); Track1Score uses our re-implementation of Eq.~(\ref{eq:score}) under exact position matching, tagging F1 in Tables~\ref{tab:main} and~\ref{tab:aux} is position-insensitive multiset F1, and CER/WER are on tagged transcripts with tags as atomic tokens.

\subsection{Main results}

Table~\ref{tab:main} compares the official baseline checkpoint (fine-tuned Whisper-medium~\cite{nvvbaseline}) with our PIR system on the same held-out set. On our held-out set, our system improves tagging F1 by $+0.27$ absolute ($0.632 \rightarrow 0.903$), cuts the tagged error rate by a factor of roughly 4 in Mandarin and 3.4 in English, and lifts the Track1Score from 36.41 to 78.48. Our submitted system is exactly this model---PIR trained on the full corpus of Section~\ref{sec:data}; its outputs come fully automatically from the pipeline of Section~\ref{ssec:infer} (no model-based post-processing, no manual correction). On the official evaluation it scores 49.82 (zh 53.02 / en 46.63), versus 33.32 (zh 34.05 / en 32.59) for the baseline; Section~\ref{ssec:analysis} analyzes the held-out--official gap.

\begin{table}[t]
  \caption{Results on our held-out set (F1 = position-insensitive multiset tagging F1; CER/WER on tagged transcripts; Track1 = Eq.~(\ref{eq:score}) averaged over both languages under exact position matching) and on the hidden official evaluation (Official). Baseline: fine-tuned Whisper-medium~\cite{nvvbaseline}.}
  \label{tab:main}
  \centering
  \footnotesize
  \setlength{\tabcolsep}{1.6pt}
  \begin{tabular}{l c c c c c c c}
    \hline
    \textbf{System} & \textbf{zh F1} & \textbf{zh CER} & \textbf{en F1} & \textbf{en WER} & \textbf{All F1} & \textbf{Track1} & \textbf{Official} \\
    \hline
    Baseline & 0.651 & 18.1\% & 0.240 & 28.0\% & 0.632 & 36.41 & 33.32 \\
    Ours (PIR) & \textbf{0.911} & \textbf{4.43\%} & \textbf{0.794} & \textbf{8.21\%} & \textbf{0.903} & \textbf{78.48} & \textbf{49.82} \\
    \hline
  \end{tabular}
\end{table}

\subsection{Ablations}
\label{ssec:ablation}

\textbf{Load-balancing loss and rebalancing.} Table~\ref{tab:aux} isolates the two training-side changes (Sections~\ref{ssec:aux} and~\ref{ssec:rebal}) under matched training budgets, evaluated on a 2,190-utterance subset of the held-out set containing all rare classes: disabling the balancing loss yields the single largest gain ($+0.17$ F1), and rebalancing stacks on top.

\begin{table}[t]
  \caption{Training-recipe ablation: full fine-tunes without PIR on the public-corpus pool, under matched training budgets; tagging F1 on a 2,190-utterance subset of the held-out set covering all rare classes.}
  \label{tab:aux}
  \centering
  \small
  \setlength{\tabcolsep}{4pt}
  \begin{tabular}{l c c}
    \hline
    \textbf{Configuration} & \textbf{F1 (rare subset)} & \textbf{en F1} \\
    \hline
    natural dist., balancing loss on & 0.652 & 0.509 \\
    natural dist., balancing loss off & 0.822 & 0.708 \\
    rebalanced, balancing loss off & \textbf{0.860} & \textbf{0.720} \\
    \hline
  \end{tabular}
\end{table}

\textbf{What did not work.} Starting from the rebalanced, balancing-loss-off configuration, scored on our held-out set: splicing ESC-50~\cite{esc50} sound effects into carrier speech helped at 150 clips ($+1.5$ Track1Score points) but hurt at 478 ($-1.2$)---spliced acoustics do not transfer; a specialist model trained on a narrowed 7-tag inventory disrupted global tag calibration ($-3$ points).

\subsection{Analysis}
\label{ssec:analysis}

\textbf{Held-out vs.\ official scores.} Our held-out set is drawn from the same public corpora as our training data and shares their domain, label conventions, and long-tailed distribution, so its scores are optimistic, in-domain estimates; the official evaluation uses unseen audio whose domain and annotation conventions may differ (see below). Held-out numbers are therefore controlled comparisons between systems, not predictions of official scores: both systems drop on the official evaluation (baseline 36.41 $\rightarrow$ 33.32, ours 78.48 $\rightarrow$ 49.82), and the drop is larger for ours, whose training data share the held-out set's provenance.

\textbf{Language gap.} Public NVV resources are overwhelmingly Mandarin ($\sim$95\% of our unified corpus), an imbalance that the 50/50 bilingual scoring punishes; on our held-out set, English positional accuracy is also weaker (English mNTD $\approx 2\times$ Mandarin). Rebalancing closes part of the gap (en F1 $0.708\!\rightarrow\!0.720$, Table~\ref{tab:aux}); the per-tag 1:1 bilingual synthesis addresses the remainder at the source (Table~\ref{tab:main}).

\textbf{Rare classes are the headroom.} Consistent with the Introduction's decomposition, the recipe changes measured in isolation (Table~\ref{tab:aux}) acted primarily on rare-tag recall and precision. Per-category F1 of our system on the held-out set still spans a wide range, from 0.997 (\texttt{[burp]}) and 0.965 (\texttt{[lipsmack]}) down to 0.755 (\texttt{[moan]}) and 0.261 (\texttt{[yawn]}); the long tail remains the dominant error source.

\textbf{Label-definition ambiguity.} Two categories are underspecified, and different resources resolve them differently. \texttt{[hum]} may denote a short grunt-like vocalization---the prevailing usage in public corpora, which our training data inherit---or sustained melodic humming close to singing; \texttt{[lipsmack]} (mouth clicking) is acoustically close to kissing sounds, which some resources file under the same label. When the convention assumed at evaluation differs from the one a system was trained on, the mismatch caps the achievable score; we flag both definitions as candidates for clarification.

\section{Conclusions}

We presented a Track~1 rich-ASR system: a bilingual tagged corpus (unified public data, per-tag-balanced synthesis, QC-filtered web crawl) plus a routing-preserving MoE recipe (balancing loss off, identity-initialized PIR) that turns pretrained expert specialization into rare-tag recall; it outscores the official baseline on both our held-out set (78.5 vs.\ 36.4) and the official evaluation (49.82 vs.\ 33.32).

\bibliographystyle{IEEEtran}
\bibliography{refs}

\end{document}